\documentclass[%
 reprint,
 amsmath,amssymb,
 aps,
prb,
]{revtex4-2}

\usepackage{graphicx}
\usepackage{dcolumn}
\usepackage{bm}
\usepackage{booktabs}
\usepackage{amsmath}
\usepackage{amssymb}
\usepackage{soul}

\usepackage{braket}

\usepackage{color}
\newcommand{\com}{\textcolor{black}}
\begin{document}


\title{Two-dimensional orbital-obstructed insulators with higher-order band topology}

\author{Olga Arroyo-Gasc\'on$^{\dagger}$}
\email{o.arroyo.gascon@usal.es}
\affiliation{Nanotechnology Group, USAL – Nanolab, University of Salamanca,
37008, Salamanca, Spain}
\author{Sergio Bravo}
\thanks{These two authors contributed equally.}
\affiliation{Departamento de F\'isica, Universidad T\'ecnica Federico Santa
Mar\'ia, Casilla 110-V, Valpara\'iso, Chile} 
\author{Mónica Pacheco}
\affiliation{Departamento de F\'isica, Universidad T\'ecnica Federico Santa
Mar\'ia, Casilla 110-V, Valpara\'iso, Chile}
\author{Leonor Chico}
\affiliation{GISC, Departamento de F\'{\i}sica de Materiales, Facultad de
Ciencias Físicas, Universidad Complutense de Madrid, E-28040 Madrid, Spain}

\date{\today}

\begin{abstract}
Obstructed atomic phases, with their realizations in systems of diverse
dimensionality, have recently arisen as one of the topological states
with greatest potential to show higher-order phenomena. 
In this work we report a special type of obstruction, known as orbital-mediated
atomic obstruction, in monolayers of materials with crystalline symmetry described
by the space group P$\bar{3}m1$. By means of a minimal tight-binding model and
first-principles calculations, we show that this obstructed phase is related
to the mismatch of the charge centers coming from the atomic limit with respect to the
centers that are obtained from a reciprocal space description. Although we find 
atomic limits that correspond with occupied atomic sites, orbital-mediated atomic
obstruction requires the presence of orbitals that have no support in real space. 
In order to demonstrate the nontrivial character of the obstruction, we confirm 
the presence of a filling anomaly for finite geometries that is directly
associated with the bulk configuration, and discuss the role of the boundary
states and their underlying mechanism.
Several material examples are presented to illustrate the ubiquity of these nontrivial
responses and, in turn, to discuss the differences related
to the particular ground state configuration.
In addition, we perform a survey of materials and elaborate a list of
candidate systems which will host this obstructed phase in monolayer form. 

\end{abstract}

\maketitle

\section{Introduction \label{Intro}}

Modern materials science has witnessed a breakthrough due to the inclusion and
adaptation of notions from topology as key tools for the analysis and discovery
of new phenomena. This approach has led to the establishment of novel phases in
materials with paradigmatic examples such as topological insulators, topological
semimetals and topological superconductors \cite{vergniory_complete_2019,
zhang_catalogue_2019,
wieder2022}. 
In recent years, this topological description has been extended to include
not only three- and two-dimensional systems that exhibit anomalous surface
or edge states, respectively, but also higher-order states. 
A $n$-th order $d$-dimensional topological
phase hosts $d-n$-dimensional states. Therefore, a new variety
of symmetry-protected states have emerged, generating great interest in
their potential applications
\cite{schindler_higher-order_2018-1,schindler_higher-order_2018,
benalcazarQuantizedElectricMultipole2017,Han2024}.

Hand in hand with these discoveries goes the theoretical description
and classification of all different topological phases. Topological 
Quantum Chemistry (TQC) theory and the associated symmetry indicators formalisms
stand out as one of the most useful tools for analyzing crystalline materials
\cite{bradlyn_topological_2017-1,Cano_EBRs_2018,po_symmetry-based_2017,
khalaf_symmetry_2018-1,po_symmetry_2020,PRX_7,Slager2012}. 
Within this setting, the band topology of any material can be analyzed in a
straightforward process and their diverse topological features can be 
identified.

In this work, we use the TQC framework to search for higher-order effects in
two-dimensional systems. 
We focus on materials for which the set of valence
bands can be described by Wannier functions, implying a well-defined
atomic limit \cite{cano_band_2021,bradlyn_topological_2017-1}.
In such scenario, the valence band set can be expressed as 
a linear combination of the so-called elementary band representations (EBRs)
\cite{bradlyn_topological_2017-1,
cano_band_2021,
Cano_EBRs_2018}. 
These EBRs correspond to the band representations that will be induced in
momentum space by different orbitals located at specific (maximal) Wyckoff
positions (WPs) in real space \cite{Cano_EBRs_2018}. 
\com{WPs are said to
be maximal if the symmetry operations that leave them invariant (dubbed
the site-symmetry group) constitute a subgroup of the space group of the
material, but are not a subgroup of any other site-symmetry groups
\cite{cano_band_2021,Cano_EBRs_2018}.}
In terms of strong topology, these systems are termed as trivial
insulators. However, since the map from real to momentum space
is not one-to-one, there can be atomic insulators whose valence band manifold 
can only be described including EBRs that are induced from WPs that do not correspond
to the atomic positions in the crystal \cite{bradlyn_topological_2017-1}.

Such materials, in which 
the valence bands contain one or more unavoidable EBRs coming
from an unoccupied WP, are denoted as obstructed atomic insulators (OAIs)
\cite{Xu_FEOAI_OOAI_2024,bradlyn_topological_2017-1}. 
A second kind of obstruction is realized when, although all the EBRs
describing the valence band come from occupied WPs, there is a mismatch
between the expected band representations that are induced by the
specific orbitals and the actual EBRs that the material contains. These insulators
are known as orbital-obstructed atomic insulators (OOAI) or unconventional atomic insulators 
\cite{xu2021_RSI_arxiv,OAI_catalysis_2022}.

The previous general definitions yield a more elaborate classification of atomic
insulators. In particular, in two-dimensional systems, corner states and
charges have been reported in OAI material candidates since 2019 
\cite{benalcazar_quantization_2019,schindler_fractional_2019-1}, allowing for the
discovery of new topological phenomena related to higher-order responses. 
Less studied are the OOAIs, which have been previously reported
in Refs. \cite{Gao_3d_OOAI_2022,Sheng_OOAI_majorana_2024,2DTQC} and associated, for
example, with obstructed edge states and superconducting corner states.  

In this work, we focus on a family of two-dimensional materials in monolayer
form, with spatial symmetry described by space group (SG) P$\bar{3}m1$
(No. 164) and with formula MX$_2$, where M can be in general a metal and X
represents a chalcogen or halogen atom. These monolayers have been
widely studied and many of them are experimentally feasible, such as 
compounds with M=Zr, Ni, Sn, Pt, Pd or Pb
\cite{
manas-valero_raman_2016,ye_synthesis_2017,tsipas_massless_2018,
xu_topological_2018,zhangExperimentalEvidenceTypeII2017,wangMonolayerPtSe2New2015,
yuanDirectGrowthVan2021,sinhaAtomicStructureDefect2020,chen2DLayeredNoble2020,
liuEpitaxialGrowthMonolayer2021}. 
Besides, several theoretical proposals have been reported in computational
databases \cite{haastrup_computational_2018,gjerding_recent_2021}.  
Our analysis uncovers an OOAI phase in this type of
materials, for which we propose a second-order bulk-boundary correspondence
associated with this obstructed state that is identified in finite
geometries by an electron filling anomaly. 

The article is organized as follows. In Section II, we review SG No. 164 from
a TQC perspective; in Section III, we propose a minimal tight-binding model
that obeys the symmetries of that SG, and discuss the emergence of a filling
anomaly, corner and edge states within that setting. In Section IV, we present
material realizations to the effects outlined in the previous sections,
introducing three representative materials that reflect the different scenarios
previously analyzed. Section V discusses the overall findings and significance
of this work.

\section{Structural and symmetry properties of space group P$\bar{3}m1$
materials}\label{sec1}

\begin{figure}[t!]
\includegraphics[width=\columnwidth,trim={10cm 4cm 16.5cm 6cm},clip]{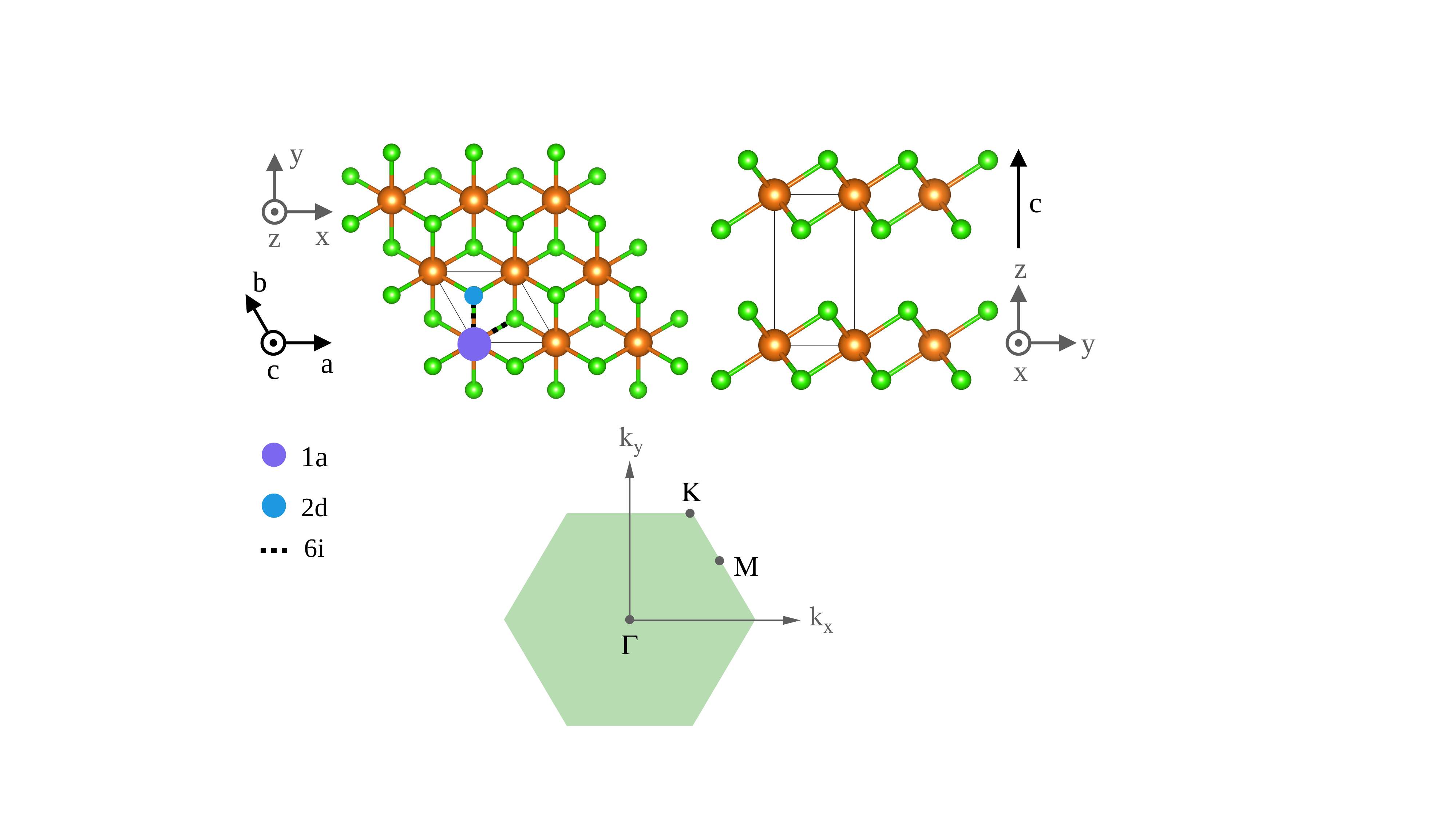}
\caption{Top panel: crystal structure of a SG No. 164 unit cell, from a top
(left panel) and side (right panel) view. The $1a$ and $2d$ WPs are indicated as
blue and orange highlighted atoms in the top left panel, and the $6i$ WP as black dashed
lines. Bottom panel: monolayer Brillouin zone.}
\label{fig1}
\end{figure}

The materials of interest in this work have a lattice structure with unit
cell depicted in Fig. \ref{fig1}, where the lattice vectors for the two
periodic directions (defined in the $xy$ plane) are expressed
as $\boldsymbol{a}=a\boldsymbol{\hat{x}}$ and $\boldsymbol{b}=-
(1/2)a\boldsymbol{\hat{x}}+(\sqrt{3}/2)a\boldsymbol{\hat{y}}$, with $a$ 
being the lattice parameter. 
The symmetries of this lattice are described by the layer version of
the SG P$\bar{3}m1$ (layer group No. 72), with twelve symmetries,
which is generated by spatial inversion, a threefold symmetry of
$2\pi/3$ about the $z$-axis perpendicular to the monolayer plane, and a
twofold rotation around an in-plane axis \cite{aroyo_bilbao_2006-1}. 
We also assume that time-reversal symmetry is preserved in all cases
considered in this work. 

The structures with formula MX$_2$ can be generically described
under this SG by the identification of the WP for each type of atom.
The M atom located at the origin of the unit cell corresponds to
the $1a$ WP, and the pair of X atoms compose the $2d$ WP. This last WP
implies that the structure has a buckled configuration 
(see Fig. \ref{fig1}). 
The occupied WPs play a key role in the symmetry description in real and
momentum space. 

As time-reversal symmetry and spatial inversion are both present, all
energy bands will be twofold degenerate along the entire Brillouin
zone (BZ). The relevant high-symmetry points of the BZ are $\Gamma$, $M$ and $K$.
Since spin-orbit coupling is also considered, double space group irreducible
representations (irreps) are considered \cite{bilbao_doubleSG}. 
\com{Note that both SG No. 164 and layer group No. 72 share the same irreps
at these high-symmetry points, so a one-to-one correspondence can be established.}
The monolayer BZ along with their HSPs is shown in Fig.\ref{fig1}. 
In this scenario, bands are at most twofold degenerate, which allows us
to study each twofold degenerate band separately.
Another important ingredient that is derived from real space symmetry 
is the type of band representation that atomic orbitals induce in
momentum space. Although we use double-valued irreps, analyzing the
symmetry features of the atomic orbitals allows to relate the \com{band
representation (BR)} with the inducing spinful orbitals at the occupied
atomic sites.

For SG No. 164 we need to consider each occupied WP separately.
For WP $1a$, the local symmetry is described by the
site-symmetry group (SSG) P$\bar{3}m1$, the complete point group
of the SG 
\cite{aroyo_bilbao_2006,aroyo_bilbao_2006-1,aroyo_crystallography_2011}. 
Under this local symmetry, the crystalline field produces atomic  level splittings 
%
that group the atomic levels in
sets with a maximal twofold degeneracy \cite{dresselhaus_group_2008}. In
addition, as inversion is present in the SSG, $p$ orbitals transform differently
from $d$ and $s$ orbitals.  
Therefore, we can identify the irreps that label
the BR for each set of orbitals, as summarized in Table \ref{TABLE_I}. 
The same process can be repeated for the $2d$ WP, whose SSG is P$3m1$.
In this case no inversion is present, and then there is no distinction 
of atomic orbitals related to spatial parity. Thus, the orbitals 
located at this WP are classified mainly by the degeneracy resulting
from the crystal field splitting. A summary of the BRs that can be
obtained from atomic orbitals located at this WP is also presented in
Table \ref{TABLE_I}. 

The data in Table I will be useful for characterizing the OOAI phase,
and is also the starting point to construct a minimal tight-binding model
that illustrates this topological phase.

\begin{table*}[]
\centering
\begin{tabular}{@{}ccccc@{}}
\toprule
WP   & Orbitals          & Irreps induced at HSPs    & EBR decomposition
\\ \midrule
$1a$   & ($s$ or $d_{z^2}) \otimes (\ket\uparrow,\ket\downarrow)$  & $\bar{\Gamma}_{8}$ - $\bar{M}_{3}\bar{M}_{4}$
- $\bar{K}_6$ & $\bar{E}_{1g}(1a)$    \\
$1a$   & ($p_{x}$,$p_{y}$) $\otimes (\ket\uparrow,\ket\downarrow)$  & $\bar{\Gamma}_{6}\bar{\Gamma}_{7} \oplus
\bar{\Gamma}_{9}$ - $2\bar{M}_{5}\bar{M}_{6}$ - 
$\bar{K}_{4}\bar{K}_{5} \oplus \bar{K}_6$ 
& $\bar{E}_{1u}(1a)\oplus {}^1\bar{E}_{u}{}^2\bar{E}_{u}(1a)$ \\
$1a$   & $p_{z}\otimes (\ket\uparrow,\ket\downarrow)$   & $\bar{\Gamma}_{9}$ - $\bar{M}_{5}\bar{M}_{6}$ - $\bar{K}_6$ &
$\bar{E}_{1u}(1a)$  \\
$1a$   & (($d_{xy}$,$d_{yz}$) or ($d_{xy}$,$d_{x^2-y^2}$)) $\otimes (\ket\uparrow,\ket\downarrow)$ &
$\bar{\Gamma}_{4}\bar{\Gamma}_{5} \oplus \bar{\Gamma}_{8}$ - 
$2\bar{M}_{3}\bar{M}_{4}$ - $\bar{K}_{4}\bar{K}_{5} \oplus \bar{K}_6$ &
$\bar{E}_{1g}(1a)\oplus{}^1\bar{E}_{g}{}^2\bar{E}_{g}(1a)$ \\
$2d$  & ($s$ or $p_{z}$ or $d_{z^2}$) $\otimes (\ket\uparrow,\ket\downarrow)$ & $\bar{\Gamma}_{8} \oplus \bar{\Gamma}_{9}$ - 
$\bar{M}_{3}\bar{M}_{4} \oplus \bar{M}_{5}\bar{M}_{6}$ - $\bar{K}_{4}\bar{K}_{5}
\oplus \bar{K}_6$ & $\bar{E}_{1}(2d)$  \\
$2d$  & (($p_{x}$,$p_{y}$) or ($d_{xy}$, $d_{yz}$) or ($d_{xy}$,$d_{x^2-y^2}$))$\otimes (\ket\uparrow,\ket\downarrow)$ &
$\bar{\Gamma}_{4}\bar{\Gamma}_{5} \oplus \bar{\Gamma}_{6}\bar{\Gamma}_{7} \oplus
\bar{\Gamma}_{8} \oplus \bar{\Gamma}_{9}$ -
& $\bar{E}_{1}(2d)\oplus {}^1\bar{E}{}^2\bar{E}(2d)$       \\ 
  & & $2\bar{M}_{3}\bar{M}_{4} \oplus 2\bar{M}_{5}\bar{M}_{6}$ - $\bar{K}_{4}\bar{K}_{5} \oplus 3\bar{K}_6$ &    \\
\bottomrule
\end{tabular}
\caption{Irreducible representations at HSPs induced from atomic orbitals
located at the $1a$ and $2d$ WP of SG No. 164. The EBR decomposition for each
induced set of irreps is presented in the fourth column. Information has been 
retrieved from BANDREP code in the Bilbao Crystallographic Server
\cite{bilbao_doubleSG,aroyo_bilbao_2006,aroyo_bilbao_2006-1,aroyo_crystallography_2011}
and adapted for a layer group configuration.}\label{TABLE_I}
\end{table*}

\section{Minimal tight-binding description of the OOAI state}

To show how the orbital-induced obstruction arises in materials
with SG No. 164, we build a tight-binding model by decorating 
the $1a$ and $2d$ WPs with atomic orbitals. The particular choice
of the orbitals has no effect on the realization of the obstruction. 
Thereby, we place a single (spinful) orbital that transforms as $\bar{E}_{1g}$
at $1a$, and one (spinful) orbital at each site of the $2d$ WP, which
transforms as $\bar{E}_1$ (see Table \ref{TABLE_I} for details on the BRs). This
yields a total of six orbitals, which in turn conform a six-band model in
reciprocal space. In the following, we use a state basis ordered as 
$\left\{ 
\left|\psi_{1a},\uparrow\right\rangle, 
\left|\psi_{1a},\downarrow\right\rangle,
\left|\psi^{A}_{2d},\uparrow\right\rangle,
\left|\psi^{A}_{2d},\downarrow\right\rangle, 
\left|\psi^{B}_{2d},\uparrow\right\rangle,
\left|\psi^{B}_{2d},\downarrow\right\rangle 
\right\}.$  The subscript of state $\psi$ denotes
the WP at which the orbital is located and, for the $2d$ states,
the superscript distinguishes between the two sites that 
compose this WP. The arrows indicate the spin character of the state, 
considering as the basis the two spin states for a quantization axis
along the $z$ direction.

By including spin-orbit coupling, we construct with the aid of the
method discussed in Ref. \cite{ZHANG_magneticTB} a Hamiltonian that respects
all the spatial transformations along with time-reversal symmetry.
The formulation in block form in momentum space will be
\begin{equation}
H =
\begin{pmatrix}
H_{1a} & H_{1a,2d} \\
H^{*}_{1a,2d} & H_{2d}
\end{pmatrix}.
\end{equation}
The $H_{1a}$ block, which describes the interaction between
the $1a$ sites, is given by (for simplicity we omit the
explicit momentum dependence in all the following functions) 
\begin{equation}
H_{1a}=    
\begin{pmatrix}
\epsilon_{1a}+F(0,r_2) & 0  \\
0 & \epsilon_{1a}+F(0,r_2)
\end{pmatrix},
\end{equation}
where $\epsilon_{1a}$ is the onsite energy for orbitals at $1a$
WP and $r_2$ is a real parameter to quantify the $1a$,$1a$ coupling.  
The interaction between the $2d$ sites, $H_{2d}$, will be
expressed as 
\begin{widetext}
\begin{equation}
H_{2d}=
\begin{pmatrix}
\epsilon_{2d}+F(r_1,r_4)) & G(r_3)& L(s_1) & 0 \\
G^{*}(r_3) & \epsilon_{2d}+F(-r_1,r_4)& 0 & L(s_1) \\
L^{*}(s_1) & 0 & \epsilon_{2d}+F(-r_1,r_4)&G(-r_3)\\
0 & L^{*}(s_1) & G^{*}(-r_3)& \epsilon_{2d}+F(r_1,r_4)
\end{pmatrix}.
\end{equation}
\end{widetext}
Here, $\epsilon_{2d}$ is the onsite energy for orbitals at the $2d$ WP
and $r_1$ parameterizes the spinless interaction, while $r_3$ and $r_4$
parameterize, respectively, the spin-flip and the spin-conserving
spin-orbit couplings. For its part, $s_1$ parameterizes the interaction
between different sites of the $2d$ WP, that is, between the top and bottom
atoms of the monolayer.
Finally, the block describing the inter-sublattice interaction,
$H_{1a,2d}$, is given by 
\begin{equation}
H_{1a,2d}=
\begin{pmatrix}
L(t_2) & \tilde{M}(t_1) & L^{*}(-t_2) & T(t_1) \\
M(t_1) & L(t_2) & \tilde{T}(t_1) & L^{*}(-t_2)
\end{pmatrix},
\end{equation}
where $t_1$ and $t_2$ parameterize, respectively, the spin-flip and
spin-conserving interactions between the orbitals located 
at $1a$ and $2d$ WPs. 
The explicit form of the functions $F,G,L,
M$, $\tilde{M}$, $T$ and $\tilde{T}$ is reported in the Appendix.
The previous model is solved numerically to obtain the band
structure for particular values of the parameters. For convenience, 
these values are chosen such that no band inversion is produced,
which means that no strong topological band is realized. This has
no impact on the results obtained. 

\begin{figure*}[ht!]
\includegraphics[width=2\columnwidth,trim={0cm 0cm 0cm 0cm},clip]{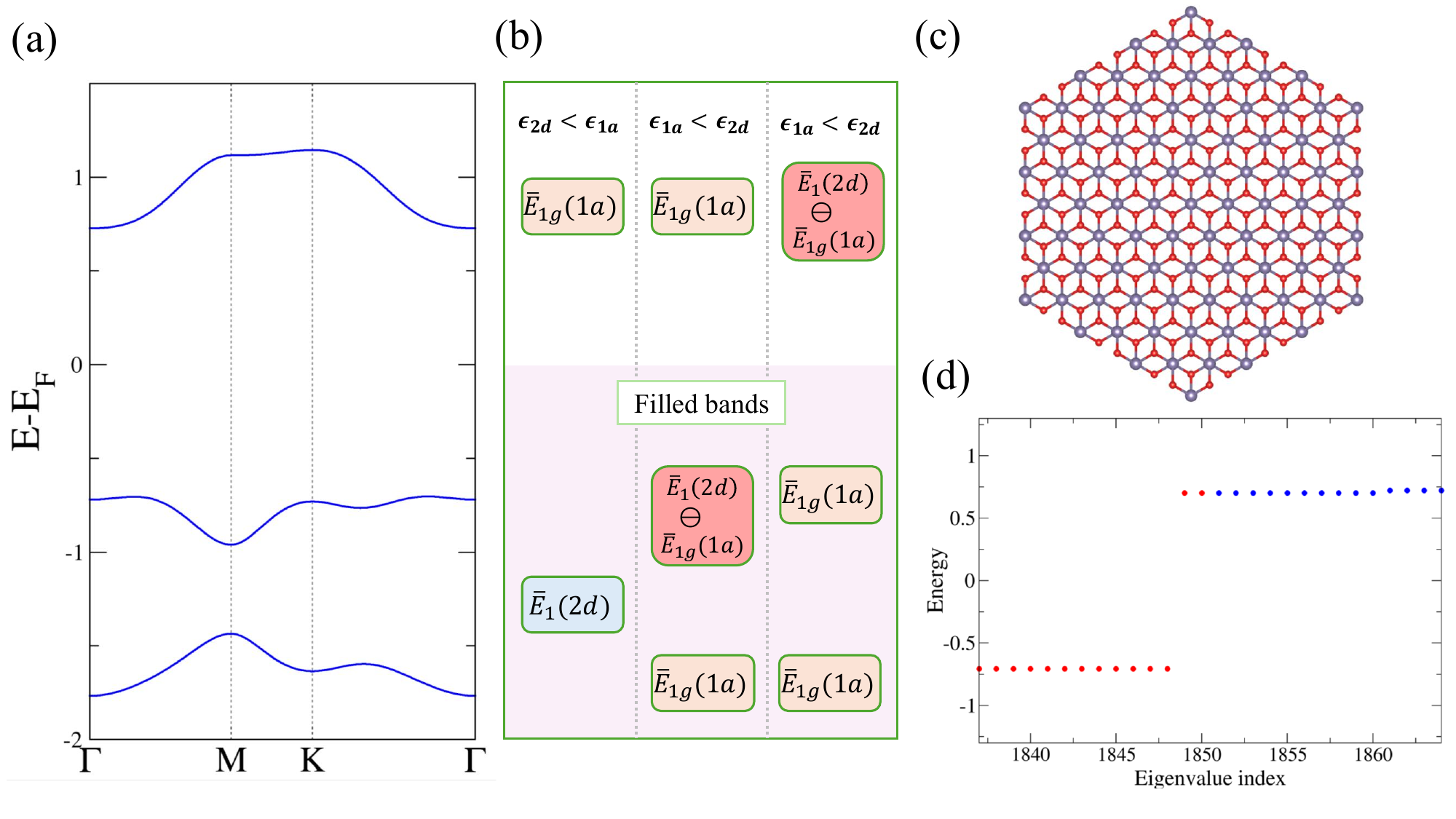}
\caption{(a) Schematic band structure for the minimal tight-binding model of Section III.
(b) Diagrammatic depiction of the occupied and unoccupied bands of the aforementioned
model, highlighting the different three regimes discussed in the text. (c) Hexagonal
flake presenting \com{P}$\bar{3}m1$ point group geometry, and (d) its corresponding eigenvalue
spectra, where the occupied states are colored red.}
\label{fig2}
\end{figure*}

In order to clarify the forthcoming discussion, which relates the
model to the material realizations, we will distinguish between
two regimes based on the relative magnitudes of the onsite energies:
A) $\epsilon_{2d} < \epsilon_{1a}$ and B) $\epsilon_{1a} < \epsilon_{2d}$.
Throughout this section, 
we will assume a bulk 
filling of $2/3$, that is, four of the six bands are completely filled. An
example of a band structure for this setting is presented in Fig. \ref{fig2}a.

\subsection{Case $\epsilon_{2d} < \epsilon_{1a}$}

As mentioned in the introduction, following TQC,
an atomic insulator can always be described as a linear
combination of EBRs induced from maximal WPs
\cite{Cano_EBRs_2018, bradlyn_topological_2017-1}. 
This implies that, in the atomic limit, we could order the bands
according to the onsite energy of the set of orbitals. 
Therefore, if we consider $\epsilon_{2d} < \epsilon_{1a}$, 
the four bands induced from the $2d$ WP will be the
lowest energy bands in the system. The same picture can 
be maintained if we turn on the hopping interactions of the
model, provided that no gap closures occur with the upper
bands. 

According to the 2/3 fixed filling we are working with, the
four low-lying bands comprise the valence bands of the model and,
for the present energy ordering, this set of bands can be
described by the EBR $\bar{E}_1(2d)$ (see leftmost panel of Fig.
\ref{fig2}b). 
This EBR is classified as \textit{decomposable} 
\cite{cano_topology_2018}, 
which means that the four-band group will be separated
in two sets of twofold degenerated bands. Depending on the 
value of parameters controlling the interaction between 
the $2d$ sites, this EBR can be expressed either as the 
combination of another EBR and a band with fragile topology
\cite{po_PhysRevLett121} or as the 
combination of two BRs with strong topology 
\cite{bradlyn_disconnected_2019}. For this particular case, the
actual decomposition does not play a role since we consider that
the four bands are filled. 
However, to simplify the discussion on edge states of
Section \ref{edge_st} 
we assume that the decomposition matches that of an EBR plus a
fragile band. 
In order to spot the obstruction, the bulk filling is studied, where
there must be 4 electrons in the unit cell. As the 
$1a$ and $2d$ WPs must be occupied, the only possibility that respects
the symmetry of the system is to have 2 electrons 
(and hence also two ions) at the $1a$ WP and one electron 
(and one ion) per site of the $2d$ WP. This configuration
is the only possibility with both WPs occupied. 
However, referring to Table \ref{TABLE_I}, 
the filled valence bands are represented by EBR $\bar{E}_1(2d)$,
in disagreement with the aforementioned only possible
arrangement in real space which involved $1a$ and $2d$ WPs.
This effect is called orbital-induced atomic obstruction, and is 
a topological bulk obstruction to connect different atomic phases
without breaking the underlying symmetries of the
system \cite{xu2021_RSI_arxiv,OAI_catalysis_2022}. 

\subsection{Case $\epsilon_{1a} < \epsilon_{2d}$}

If the orbitals at $1a$ are lower in energy than the $2d$ orbitals, 
the bands induced from the $1a$ WP will be completely filled and
the decomposable EBR induced from the $2d$ WP will be half-filled.
Following the same procedure as in case A,  
and setting the last filled band as a fragile band, we end up with a group of
four filled bands that is again described by EBR $\bar{E}_1(2d)$, as
illustrated in the middle panel of Fig. \ref{fig2}b.  
Therefore, an orbital obstruction is again achieved. This scenario,
with a half-filled EBR, has been previously analyzed in Ref. 
\cite{Sheng_OOAI_majorana_2024}.

There is one more instance within this regime of onsite energy
ordering, which is sketched in the rightmost panel of Fig. 
\ref{fig2}b. If the last filled band is not a fragile band but an
EBR, the valence bands of the model turn out to be described
by a sum of two EBRs coming from the $1a$ WP, namely, $2\bar{E}_{1g}(1a)$.
This scenario also yields an orbital obstruction, since now 
all charge centers are predicted to be located at $1a$ in 
momentum space, which is again not consistent with the atomic
positions in real space.

Hence, we have presented three kinds of obstructions in this SG, which
will show a non-trivial behavior when considering systems of reduced
dimensionality. 

\subsection{Filling anomaly induced from an OOAI phase}

The bulk obstruction that is present in these systems
can yield anomalous responses in a finite geometry setting.
The $N$ sites of such a finite system can be decomposed into a sum
of $n_{1a}$ sites that are related to the bulk $1a$ WP, and
$n_{2d}$ pairs of sites that can be mapped to the bulk $2d$ WP.
Considering the filling of the bulk (4 electrons per unit cell in
the model of Section III), we observe that the expected filling for
the finite geometry to be gapped is $4n_{2d}$. 

On the other hand, if we impose charge
neutrality and a symmetric configuration, we end up with a 
total electronic charge of $2n_{1a}+2n_{2d}$. In an open system 
these two values can differ, implying that the
system cannot be in a state that is simultaneously gapped,
symmetric and charge neutral, and a filling anomaly arises
\cite{benalcazar_electric,benalcazar_quantization_2019,
khalaf_boundary-obstructed_2021,schindler_fractional_2019-1}.
This reasoning can be generalized to include arbitrary bulk
fillings $\nu$ and associated $1a$ and $2d$ site-fillings
$\alpha_{1a}$ and $\alpha_{2d}$, respectively; therefore,
a filling anomaly appears if
$\nu n_{2d} \neq \alpha_{1a}n_{1a}+\alpha_{2d}n_{2d}$.
This result is an indicator that relies on finite geometry quantities.

However, the obstruction is a bulk property, and in consequence,
the filling anomaly should be predictable from bulk
features. In the previously described minimal model,  
for the case A of onsite energy ordering (see leftmost panel of
Fig. \ref{fig2}b), the bulk
characterization of the obstruction is very straightforward.
Since the complete set of valence bands at 2/3 filling can be
described by the EBR $\bar{E}_1(2d)$, 
the (four) electron charge centers should be
located at this WP. 
In contrast, in a symmetric configuration, two charge centers are
expected to appear in position $1a$, and the rest in position $2d$.
It follows that there is a mismatch of two electrons in the system,
which quantifies the anomaly. 
Therefore, for a generic system the filling anomaly can be
predicted through the determination of the number of
obstructed states in the system, which is obtained by comparing
the predicted charge centers attending to momentum space
information against those obtained by analysis of the real-space
orbital content.

To illustrate this anomalous behavior, we consider a particular
finite configuration corresponding to a 
hexagonal flake, which preserves point group $\bar{3}m1$ and is 
presented in Fig. \ref{fig2}c. 
Using the minimal model and considering case A with the 
same parameters as those of Fig. \ref{fig2}a, we compute the energy
spectra for a hexagonal flake with $N=1387$ sites. The result is depicted
in Fig. \ref{fig2}d for the range of energies where the bulk gap
appears; the filled states are highlighted in red. The
filling is calculated imposing charge neutrality and a symmetric
system, which results in $n_{1a}=463$ and
$n_{2d}=462$, and a filling of $2\cdot463+2\cdot462=1850$.
This differs from the filling needed in order to have a gapped
configuration, which is $4 n_{2d}=4\cdot 462 = 1848$. In agreement with the
foregoing analysis of the bulk properties, there are two states that cause
the filling anomaly.

In the $\epsilon_{1a} < \epsilon_{2d}$ regime (case B),
when the EBR induced by $2d$ is partially filled by a fragile band,
we also find a similar filling anomaly. Finally, in the case
illustrated by the rightmost panel of Fig. \ref{fig2}b, 
the method to quantify the anomaly outlined above is still
applicable, yielding a filling anomaly of
two. This last case has not been encountered in the family 
of materials scoped in this work. 
In Section \ref{Materials}, examples for cases A
($\epsilon_{2d} < \epsilon_{1a}$) and 
B ($\epsilon_{1a} < \epsilon_{2d}$) with the fragile band in 
the filled set will be presented. 

Notably, it is worth mentioning that this anomaly does not
require the presence of corner states since, albeit it is one
of the properties with which second-order topology and other
topological phases have been characterized in previous works
\cite{benalcazar_quantization_2019,schindler_fractional_2019-1,PhysRevB.103.195145,wang_higher-order_2019,kooi_bulk-corner_2021,nuñez_2024,costa_connecting_2023,qian_c_2022,zeng_multiorbital_2021,liu_magnetic_2023,PhysRevMaterials.8.044203,PhysRevB.105.L201301}, it
is not really a necessary condition for its presence \cite{Jung_BBcorr_2021}.
In the following, we will discuss the role of the boundary states
in terms of our minimal model.

\subsection{Edge and corner states} \label{edge_st}

\begin{figure*}[t!]
\includegraphics[width=2\columnwidth,clip]{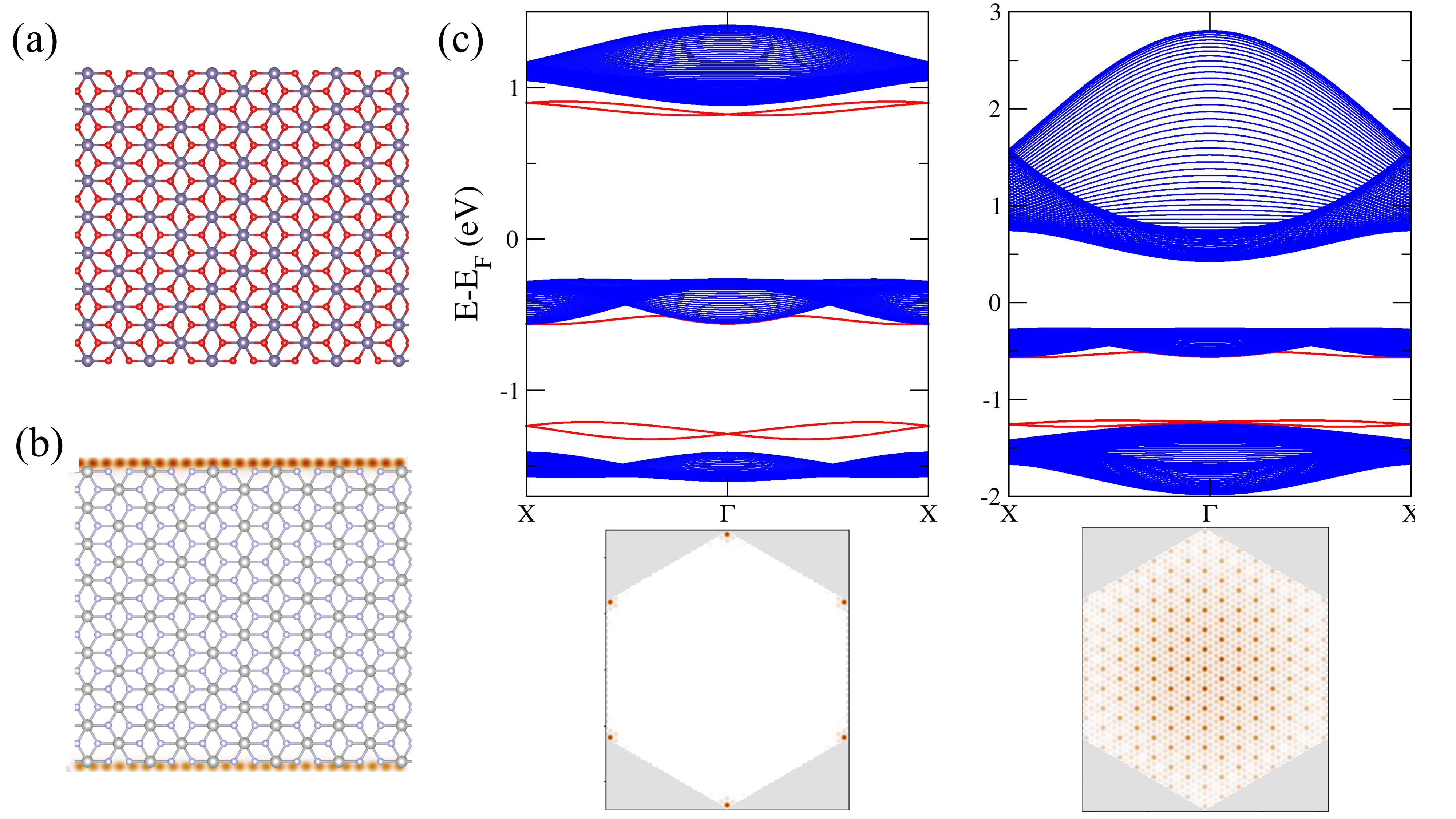}
\caption{(a) Finite-width ribbon geometry, with an edge termination analogous
to that of the hexagonal flake of Fig. \ref{fig2}c. Edges run horizontally.  
(b) Local density of states for the
ribbon geometry, computed at 0.85 eV, where edge states are expected. 
(c) Top left panel: band structure for 
$\epsilon_{1a} = 0.2$, $\epsilon_{2d} = -0.5$, $t_1 = 0.2$, $t_2 = 0.3$,
$r_1 = -0.01$, $r_2 = 0.08$, $r_3 = 0.05$, $r_4 = -0.04$, $s_1 = 0.08$.
Bottom left panel: corresponding charge density of a hexagonal flake with the same set of
parameters. Top right panel: band structure for the same
parameters, excepting   
that now $r_2 = 0.36$. Bottom right panel: corresponding charge density of
a hexagonal flake with the set of parameters for the model as in the top right panel.}
\label{fig3}
\end{figure*}

The presented minimal model can also describe the most salient 
qualitative properties of the edge states found in SG No. 164
monolayers. For the sake of brevity,
we will focus on one type of border, since we have
numerically checked that the other possible configurations do
not substantially alter the conclusions.
We build a finite ribbon, depicted in Fig. \ref{fig3}a;
in this case, the edges are similar to the boundaries of the flake
structure of Fig. \ref{fig2}c.
To explore the enabling conditions for the emergence of edge states,
we numerically survey the ribbon energy spectra and local density 
of states at certain energy values as a function of the parameters
of the model. We first describe the results for case A 
($\epsilon_{2d} < \epsilon_{1a}$) ordering, and then comment on the
reverse regime B. 

In Fig. \ref{fig3}c we show two exemplary cases that describe the 
two major possibilities for edge physics. 
In general, we find that at the tight-binding level, the spin-orbit 
interaction with spin-flip character between different WPs is key for the
generation of isolated-in-energy edge states. 
The left panel of Fig. \ref{fig3}c illustrates a case when this
interaction is strong enough to generate an isolated set of edge
states that are located inside the energy gap. For this regime of
onsite energies, the evolution of the edge states with respect to
the magnitude of the hopping parameter for the above-mentioned
interaction indicates that the edge states detach from the upper bands.
The origin of these states can be traced to the hybridization process
between the $1a$ sublattice and the $2d$ sublattice. 

If in a first instance we consider no inter-sublattice interactions, 
the $2d$ sublattice will present edge states within the lower gap 
which are due to the decomposable nature of the EBR. 
These states are gapped edge states which are inherited from a strong
topology phase that arises in a high-symmetry planar structure. Once buckling
is present, the gapless states become gapped due to the
breaking of the in-plane mirror symmetry.
However, they retain some of their spatial localization and, independently if they
are gapless or gapped, they can hybridize with the upper bands in the
model.
Once we turn on sublattice interaction, the edge states produced by the
$2d$ sublattice are induced by proximity into the $1a$ sublattice, endowing
the upper bands with edge states if the interaction is sufficiently strong. 
This picture is supported by the local density of the wavefunction, as shown
in Fig. \ref{fig3}b, where the spatial localization is evident. This calculation
is performed at the energy value where the upper edge states reside (0.85
eV, from the left panel of Fig. \ref{fig3}c).

The spin-orbit interaction that produces the isolated edge states has to 
compete with other interactions in the system that tend to 
mask the emergence of edge states. The main competing feature
is the energy width of the upper bands. This is mainly affected by 
the intra-sublattice interaction ($r_2$) between the $1a$ sites, which
greatly affects band dispersion; if this interaction is strong enough, 
no edge states will arise. 
\com{This is also related with the fact that the edge states induced by
spin-orbit coupling in this type or energy ordering stem from the conduction
states, and as such, the edge states will be in general near the bottom
of the conduction bands. Therefore, the interaction described by the
$r_2$ parameter, which affects the dispersion of the conduction bands, will 
also affect the apparition of the edge states}. This is the case depicted in the right panel
of Fig. \ref{fig3}c (see the value of the model parameters in the caption
of this figure). 
In this scenario, the edge states are completely inside the bulk 
bands and therefore are not in general protected to be in-gap. 

The above discussion illustrates that the edge state physics is
independent of the atomic obstruction, and is produced by 
hybridization and proximity effects. 
This way, by tuning the inter- and intra-sublattice interactions, the presence of 
edge states can be in principle controlled in a ribbon
geometry. Interestingly, the emergence of in-gap states in the ribbon
geometry is directly linked with the occurrence of in-gap corner
states: there is a non-protected edge-to-corner correspondence
that can be useful to diagnose the presence of corner states solely
by the edge spectrum. To further demonstrate this relation, we resort to the finite
hexagonal geometry in Fig. \ref{fig2}c, for
the cases exposed in Fig. \ref{fig3}c, 
and compute the charge density for the states that form the
filling anomaly. 
The results are depicted at the lower insets of each panel in 
Fig \ref{fig3}c, where the presence of corner states is directly
linked to the edge states of the ribbon. 
This further demonstrates that the filling anomaly does not need for
the presence of corner states to be established, since in both cases 
a filling anomaly appears. 
 
When the onsite energy ordering is inverted in case B ($\epsilon_{1a}<\epsilon_{2d}$),
the energy gap active at a bulk filling of 2/3 is the
gap coming from the decomposable $2d$ EBR. 
The edge states in this case correspond to states stemming directly from the gapped 
decomposable EBR. These states also hybridize with the now lower in
energy $1a$-induced bands, and then the same proximity-induced effect
is produced. However, these induced states are now inaccessible due to
the filling. 

\section{Material examples} \label{Materials}


\subsection*{MgCl$_2$}

The first selected material is monolayer MgCl$_2$, where the Mg atoms
sit at the $1a$ WP and the Cl atoms form the $2d$ WP. The
electronic band structure from first-principles calculations is presented in
Fig. \ref{fig4}a, where one of the advantages of this material becomes
readily apparent; it has a large bulk band gap (about 5.0 eV for the level
of theory used in the first-principles calculations). This is beneficial
as it increases the likelihood of observing corner states more clearly. 

To probe this, we compute the energy spectrum of a finite hexagonal geometry with the
same configuration as in Fig. \ref{fig2}c. 
The spectrum is plotted in Fig. \ref{fig4}g for the 
range of energies where the bulk band gap locates, with colored states
according to the electronic filling. 
The large bulk gap also results in a sizable gap in the finite
geometry, even for small structures, which enables 
the corner states to be directly spotted in the spectrum. 
As expected from time-reversal symmetry and point group symmetry, there are
twelve in-gap states 
in a narrow energy range, and ideally for larger structures, the
states should be degenerated in energy.
The spatial localization of these corner states can be explored with the
local density of states 
as sketched in Fig. \ref{fig4}d.
Turning to the filling anomaly, the spectrum of Fig. \ref{fig4}g 
shows a gapless system with the electronic filling inside the corner
states; there must be two filled states within this set and thus the
system realizes an obstructed phase, with two obstructed states. 

In the following, we analyze the origin of the topological obstruction
and the filling anomaly from the bulk band structure perspective.
We set as a reference the atomic electronic
configurations for each atom in the unit cell: 
$2s^2 2p^6 3s^2$ for Mg and $3s^2 3p^5$ for Cl.
Next, we perform a first-principles ground state calculation to
obtain the irreps at the three HSPs in momentum space 
(see Fig. \ref{fig1}). 
We present the details of this calculation in the Appendix.
If we follow the practice of symmetry indicators, we can form a
symmetry data vector $\boldsymbol{B}$ that summarizes the multiplicities
of each irrep for a particular BR \cite{elcoro_application_2020}. We fix
the order of the irreps as 
$(\bar{\Gamma}_{4}\bar{\Gamma}_{5},\bar{\Gamma}_{6}\bar{\Gamma}_{7},
\bar{\Gamma}_{8},\bar{\Gamma}_{9},\bar{M}_3\bar{M}_4,\bar{M}_5\bar{M}_6,
\bar{K}_4\bar{K}_5,\bar{K}_6)$. For MgCl$_2$ the vector
$\boldsymbol{B}_{MgCl_{2}}$ is 
\begin{equation}
  \boldsymbol{B}_{MgCl_{2}} = (1,2,4,5,5,7,4,8).    
\end{equation}
This representation allows us to explore the possible decompositions into
sums of EBRs. As it is well known, there can exist several valid
decompositions for one band representation.
The two starting ingredients are
the vector $\boldsymbol{B}$ and a matrix that summarizes the irrep
content for each EBR that the space group has, the so-called EBR matrix,
which is denoted simply as $EBR$. 
The explicit form of this matrix for SG P$\bar{3}m1$ is presented in
the Appendix. 
Thereby, the existence of one o more solutions, which we denote as
$\boldsymbol{X}$, can be compactly formulated as the outcome of the
following matrix equation
\begin{equation}
    EBR \cdot \boldsymbol{X} = \boldsymbol{B}.
\end{equation}
In order to explore these solutions, we use the Smith 
decomposition theory as developed in Ref. \cite{elcoro_application_2020}, that
allows to compute the vectors $\boldsymbol{X}$. 
The procedure can be briefly described as follows. The EBR matrix,
which is in general a rectangular $n \times m$ matrix,
can be Smith-decomposed in the form $\Delta = L\cdot EBR \cdot R$, 
where $L$ ($m \times m$) and $R$ ($n \times n$) are unimodular
matrices while $\Delta$ is a $n \times m$ matrix with nonzero entries
only for the $\Delta_{ii}$ elements, with $i<r$, $r$ being the rank of
the EBR matrix. Using these new defined matrices the original 
problem for $\boldsymbol{X}$ can be cast as 
\begin{equation}
    \Delta \cdot Y = C,
\end{equation}
where $Y = R^{-1}X$ and $C=LB$. Then, the solution is now expressed
as 
\begin{equation}
    X=RY,
\end{equation}
with $Y$ having the explicit form
\begin{equation}
    Y = (c_1/\Delta_{11},c_1/\Delta_{11},...,
    c_r/\Delta_{rr},y_{1},...,y_{N_{EBR}-1})^T,
\end{equation}
where the parameters $y_{1},...,y_{N_{EBR}-1}$ are integers which are
free to be adjusted to obtain different solutions. 

Applying this procedure to the vector $\boldsymbol{B}_{MgCl_{2}}$, the
generic solution for this material will be 
\begin{multline}
        \boldsymbol{X} = (2-y_{1}-y_{2},3-y_{1}-y_{2},
                      5-2y_{1}-2y_{2},\\ 
                      6-2y_{1}-2y_{2},
                      -1+y_{1},-1+2y_{1},y_{2},y_{2}).
\end{multline}
Next, we restrict $\boldsymbol{X}_{MgCl_{2}}$ to correspond
with the adequate atomic insulator, 
imposing all
components to be positive integers or zero. 
In addition, we further limit the range of solutions discarding the
cases $y_{2}\neq 0$ since they imply that the charge centers
are localized at unoccupied sites, the $3e$ WP. As we are
interested in the OOAI phase, we only keep the solutions that
involve EBRs induced from occupied WPs. 
Therefore, the final set 
of admissible solutions is 
\begin{align}
    \boldsymbol{X}^{A}_{MgCl_{2}} &= (1,2,3,4,0,1,0,0),\\
    \boldsymbol{X}^{B}_{MgCl_{2}} &= (0,1,1,2,1,3,0,0).
\end{align}
Hence, more than one decomposition will represent 
the same insulating state. 

With the aim to characterize the orbital obstruction, we will
introduce a more detailed classification for the insulating solutions. 
This is motivated by the concept
of \textit{movable} band representations and WPs
\cite{khalaf_boundary-obstructed_2021}. This type of
representation is defined as a combination of irreps that can be
induced from a set of maximal WPs and can also be
induced from the WPs that connect those maximal WPs in real space. 
Thus, for SG No. 164, we have two maximal WPs of interest,
$1a$ and $2d$, and there is one non-maximal WP connecting them,
the $6i$ WP, as depicted in the top left panel of Fig. \ref{fig1}. 
The set of irreps that can be interchangeably induced by these three WPs
is given by 
\begin{align}
\begin{split}
     \Gamma &: \hspace{5mm} \bar{\Gamma}_{4}\bar{\Gamma}_{5},
     \bar{\Gamma}_{6}\bar{\Gamma}_{7},2\bar{\Gamma}_{8},
     2\bar{\Gamma}_{9}.\\
     M  &: \hspace{5mm}  3\bar{M}_3\bar{M}_4,3\bar{M}_{5}\bar{M}_{6}.\\
     K  &: \hspace{5mm}  2\bar{K}_4\bar{K}_5,4\bar{K}_{6}.
\end{split}
\label{EBR_6i}
\end{align}
For this band representation, 
the charge centers can be adiabatically transported along the crystal
structure respecting all the symmetries, thus the name movable BR
\cite{khalaf_boundary-obstructed_2021}. 
It follows that this set cannot generate pinned BRs, which 
is a requisite for an obstructed phase to appear. \com{In this context, a BR is said to be pinned if it is not movable.}
Therefore, in the subsequent analysis, this movable set will
be subtracted from the insulator solutions obtained before. 
For the structures of interest, the movable atomic insulator
(MAI) state has two equivalent vector representations, one per
each maximal WP: 
\begin{align}
\begin{split}
    \boldsymbol{X}^{1a}_{MAI} &= (1,1,2,2,0,0,0,0).\\
    \boldsymbol{X}^{2d}_{MAI} &= (0,0,0,0,1,2,0,0).
\end{split}
\label{EBR_MAI}
\end{align}
Now we compute the pinned solutions 
\begin{align*}
    &\boldsymbol{X}^{A}_{MgCl_{2}}-\boldsymbol{X}^{1a}_{MAI},\\ 
    &\boldsymbol{X}^{B}_{MgCl_{2}}-\boldsymbol{X}^{2d}_{MAI}, 
\end{align*}
and obtain the same solution in both cases, which we denote as
$ \boldsymbol{X}^{O}_{MgCl_{2}}$, such that
\begin{equation}
    \boldsymbol{X}^{O}_{MgCl_{2}} = (0,1,1,2,0,1,0,0).
\end{equation}
If this solution is inspected, we find that there
are EBRs pinned to $1a$ and to $2d$ WPs. Nevertheless, all the EBRs
from $1a$ have atomic support in real space, while the single 
EBR induced from $2d$ has partial support in real space. Thus, 
the OOAI phase in this material can be precisely
identified as a mismatch of one EBR pinned at $2d$. 

To characterize the anomaly, the number of electrons in the 
obstructed EBR must be quantified. For MgCl$_2$ there exist one 
electron in the $2d$ EBR for each Cl atom, which means that there
is effectively only one mismatched state per Cl atom. Thus, the
filling anomaly will consist of two obstructed states. This is
in exact agreement with the energy spectrum of the 
finite geometry as presented in Fig. \ref{fig4}g.

\subsection*{ZrS$_2$} 
The next material to be studied is monolayer ZrS$_2$. The motivation to 
present this structure is twofold. First, its energy gap is 
smaller than that of MgCl$_2$, which allows to compare
the corresponding effect on the corner states. From the bulk band structure
calculation, as depicted in Fig. \ref{fig4}b, the 
gap is of about 1.4 eV, which also yields a smaller energy gap in the finite 
structure. This is shown in Fig. \ref{fig4}h, where the energy spectrum of the
flake geometry is presented. As expected, a filling anomaly can be observed. 
We note that in this case a different corner termination has been used, in order
to probe for the presence of the filling anomaly and corner states in a different
setting. The atomic configuration of the corner can be visualized
in Fig. \ref{fig4}e. 

From the energy spectrum in Fig. \ref{fig4}h, \com{the corner states have
merged with the upper states linked to the conduction band of the bulk. In fact, the band gap of ZrS$_2$ is over five times smaller than that of MgCl$_2$, and has the lowest band gap among the compounds presented in Fig. \ref{fig4}. For the sake of consistency with the rest of eigenvalue spectra presented in Fig. \ref{fig4}, the corner states are highlighted in a similar way as for the rest of panels.} 
The LDOS at an energy range around the neutrality point
is sketched in Fig. \ref{fig4}e, showing that  
the corner localization still survives; 
as explained by the model, the
support of these states comes almost exclusively from the $1a$ sites near the 
corners. 
This allows us to associate the obstructed phase of this
material with the case where the energy ordering of the atomic limit
implies that there are one or more EBRs coming from the $2d$ WP that are
completely filled. 

The second reason for highlighting this material is shown in
Fig. \ref{fig4}h: the filling anomaly has now a different
value involving four obstructed states. 
To explain this value, we follow the same procedure as in MgCl$_2$, and
only report here the final results. 
We first consider the atomic valence electronic
configurations, which we set to be $3d^{10} 4s^2 4p^6 4d^2 5s^2$ for Zr
and $3s^2 3p^4$ for S. On the other hand, from the first-principles 
band structure calculation, the BR vector is
\begin{equation}
  \boldsymbol{B}_{ZrS_{2}} = (1, 2, 4, 5, 5, 7, 4, 8).    
\end{equation}
The valid solutions are
\begin{align}
    \boldsymbol{X}^{A}_{ZrS_{2}} &= (1, 2, 3, 4, 0, 1, 0, 0),\\
    \boldsymbol{X}^{B}_{ZrS_{2}} &= (0, 1, 1, 2, 1, 3, 0, 0).
\end{align}
After subtracting the MAI representations given by Eq. \ref{EBR_MAI}, the two 
solutions yield the same pinned configuration:
\begin{equation}
    \boldsymbol{X}^{O}_{ZrS_{2}} = (0,1,1,2,0,1,0,0).
\end{equation}
Therefore, the pinned EBRs
from the $1a$ WP have atomic support, while the pinned EBR from
$2d$ WP has no support at all. 
This is different from the MgCl$_2$ case, since here both electrons
of each S atom in the unit
cell realize an obstruction. In consequence, a total of four
electrons are mismatched, which traduces in a four-state filling
anomaly.
This is in agreement with the result obtained in the finite geometry
calculation, and is a realization of a bulk-boundary correspondence.

\subsection*{SnS$_2$} 
\begin{figure*}[t!]
\includegraphics[width=\textwidth,trim={0cm 8.5cm 0cm 0cm},clip]{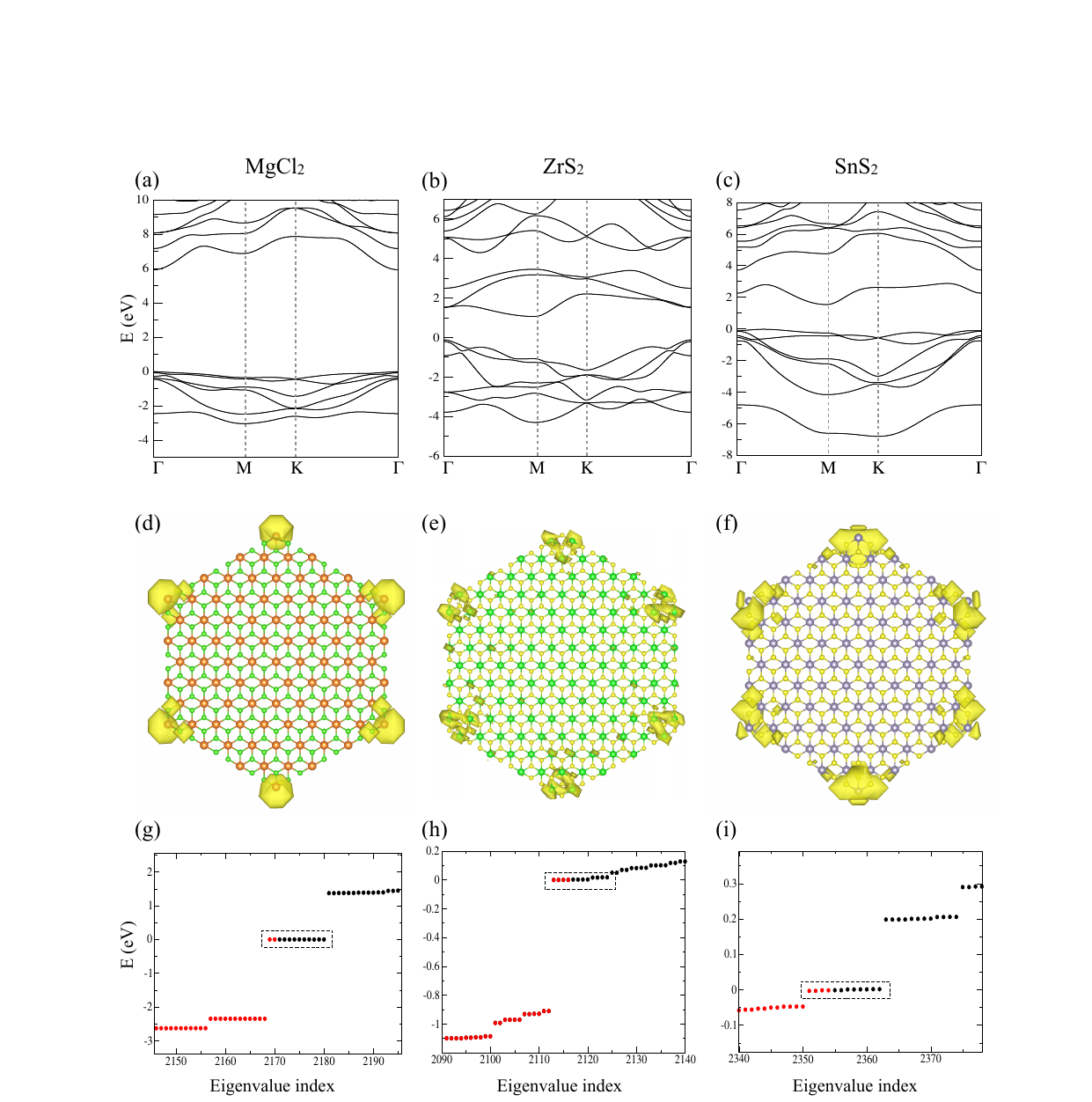}
\caption{Top panels (a - c): band structures of the three highlighted MX$_2$
monolayers belonging to SG No. 164. Middle panels (d - f): LDOS for MgCl$_2$,
ZrS$_2$ and SnS$_2$ hexagonal flakes. Bottom panels (g - i): corresponding
eigenvalue spectra of the hexagonal flakes depicted in the previous panels. \com{The corner states are highlighted by enclosing them in rectangular regions near the neutrality point.}}
\label{fig4}
\end{figure*}

Finally, we present a material showcasing the situation
where the EBR at the Fermi level is induced by the $2d$ position, in order  
to discuss the differences with respect to 
the two preceding materials. 
For this purpose, monolayer SnS$_2$ is selected, whose electronic 
bulk band structure is plotted in Fig \ref{fig4}c. The band gap is
of about 1.9 eV. 

As can be deduced from what the model asserts, this case with
half-filled EBR presents boundary responses that are coming from
the valence band. This means that the filling anomaly will manifest
in an under-filling with respect to the filling necessary for a 
gapped configuration. To verify this statement, the energy spectrum
is computed and presented in Fig. \ref{fig4}i, which is in agreement with
the physical picture provided by the model. 
There is a deficit of eight electrons 
at the neutrality point.
In our case, the filling anomaly is only well defined modulo 12, since the band
representation corresponding to the MAI (see Eq. \ref{EBR_MAI}) 
involves 12 electrons and, by definition, they cannot be pinned to any WP
in real space. 
Thus, the anomaly of SnS$_2$ will be equivalent to an obstruction of
4 electrons ($-8$ mod 12).

To check this, we apply the Smith decomposition procedure as in the previous cases. For Sn we use the valence electronic
configuration $5s^2 5p^2$ and for S we employ  
$3s^2 3p^4$. The symmetry data vector is
\begin{equation}
  \boldsymbol{B}_{SnS_{2}} = (3,1,6,3,9,4,5,8).    
\end{equation}
And the valid solutions are
\begin{align}
    \boldsymbol{X}^{A}_{SnS_{2}} &= (3, 1, 5, 2, 0, 1, 0, 0),\\
    \boldsymbol{X}^{B}_{SnS_{2}} &= (2, 0, 3, 0, 1, 3, 0, 0).
\end{align}
Subtracting the appropriate MAI representation given by Eq. \ref{EBR_MAI} to
the above solutions,
the outcome is the same for both cases and is given by 
\begin{equation}
    \boldsymbol{X}^{O}_{SnS_{2}} = (2,0,3,0,0,1,0,0).
\end{equation}
It is straightforward to see that there is a pinned EBR coming
from the $2d$ WP, which does not have support in real space. This is
due to the fact that the real space atomic occupation locates the 
electrons at the external shell of the Sn atoms ($5s^2 5p^2$). 
Hence, 
the mismatch will be of two electrons per S atom, resulting in an overall
filling anomaly of
four, which coincides with the finite geometry
calculation.

The character of the boundary states at the energy range where the 
anomaly is realized can be explored in terms of the corresponding LDOS 
for the finite geometry. This information is presented in
Fig. \ref{fig4}f, illustrating that the states that
contribute to the vicinity of the neutrality point come from the $2d$
sites.
This is expected, because the edge states are induced by the
upper bands of the system (see Section \ref{edge_st}), 
and these upper bands for SnS$_2$ are induced from the $2d$ WP. In
addition, we find here that the corner and edge states mix and yield a
LDOS contribution that extends along a portion of the flake boundary. 
These states can be directly related to bands with fragile topology, which
are feasible in this SG, and that for this monolayer occur right at
the top of the valence manifold \cite{ArroyoGascon2023}. 

\begin{table}[]
\centering
\begin{tabular}{@{}cl@{}}
\toprule
\multicolumn{1}{l}{Filling anomaly value} & \multicolumn{1}{c}{Material} \\ \midrule
2 & \begin{tabular}[c]{@{}l@{}}
CdCl$_2$, CdBr$_2$, CdI$_2$, CdF$_2$\\
RuCl$_2$, RuBr$_2$, RuI$_2$\\
ZnCl$_2$, ZnBr$_2$, ZnI$_2$\\ 
MgCl$_2$, MgBr$_2$, MgI$_2$\\ 
CaBr$_2$, CaI$_2$\\ 
SrBr$_2$, SrI$_2$\\ 
OsBr$_2$, OsI$_2$\\ 
BaBr$_2$, BaI$_2$\\
PbBr$_2$, PbI$_2$\\
ZrF$_2$, 
HgF$_2$, 
PdCl$_2$, 
SnI$_2$, 
GeI$_2$
\end{tabular} \\ \midrule
4  & \begin{tabular}[c]{@{}l@{}} 
PdO$_2$, PdS$_2$, PdSe$_2$\\
HfO$_2$, HfS$_2$, HfSe$_2$\\
PtO$_2$, PtS$_2$, PtSe$_2$ \\
ZrO$_2$, ZrS$_2$, ZrSe$_2$ \\
NiO$_2$, NiS$_2$\\
PbO$_2$, PbS$_2$\\
SnO$_2$, SnS$_2$\\ 
GeO$_2$, GeS$_2$\\
SiSe$_2$\end{tabular}                                                                 
\end{tabular}
\caption{Filling anomaly value for the studied materials. 
Recall that this value is well
defined only modulo 12. }
\label{TABLE_II}
\end{table}

\section{Discussion}

In this work we have identified a topological phase realized as an
obstructed atomic limit due to the orbital character and energy
ordering for systems within SG No. 164.
Using the TQC formalism, we have provided a detailed
characterization of the mechanism that allows the emergence of this
orbital obstruction. 
In addition, we have discussed the effect of this phenomenon in 
a finite geometry, leading to a filling anomaly.  
Remarkably, this filling anomaly is robust and does not depend on
the presence of edge states in the system.  
Moreover, using a complementary minimal six-band model, the
conditions for the occurrence of edge states in both ribbon and
finite systems have been discussed. 

The exemplary materials presented in the previous section constitute
realistic platforms for exploring the consequences of topological orbital 
obstruction. As this phenomena depends on the orbital character of
the constituents and on the energetic ordering that is realized in 
each material, this study can be readily extended to encompass the
entire family of MX$_2$-type materials within SG No. 164.
We present a summary of the calculations of the type of bulk orbital
obstruction for a set of monolayers extracted from the database in 
\cite{haastrup_computational_2018,gjerding_recent_2021} in Table II, where  
the magnitude of the filling anomaly is
reported for each material. \com{Moreover, since our analysis is based on
general TQC algorithms, it can also be extended to other SG No. 164
materials, as well as to other space groups.}

Our findings clarify and complement previous
results on nontrivial topological properties. In particular, comparing
with Ref. \cite{PhysRevMaterials.8.044203}, we found a different mechanism
for the higher-order response of the hexagonal finite geometries, 
which best accounts for all possible results for the family of materials
with SG No. 164.
Thus, if for example, the theory of symmetry-indicator
invariants is applied to materials like MgCl$_2$, we encounter an
incorrect value for the filling anomaly, since this invariant does not
account for the orbital obstruction as the main responsible of the 
nontrivial phase for halogen-based monolayers.  
On the other hand, considering Ref. \cite{Sheng_OOAI_majorana_2024},
we find similar conclusions related to materials with a half-filled EBR
coming from the $2d$ WP, for instance SnS$_2$, which is also reported
in the aforementioned work.  
The novelty of our analysis is that we identify a new effect related to these
orbital-obstructed phases: a filling anomaly. We also extend the nontrivial
behavior to cases that were previously overlooked, 
such as materials with completely filled EBRs coming from an atomic limit
related to the $2d$ WP, which include several monolayer transition metal
dichalcogenides. 

\com{Regarding the formal interpretation of the orbital obstructed phases discussed
in this work, following Ref. \cite{song_fragile_2020}, the solutions that 
constitute an OOAI are affine monoids. In particular, the affine monoids of interest have 
a trivial Hilbert basis and can be defined as the monoids that arise when from 
the complete EBR monoid we substract the monoid of 
the EBRs with electronic support. 
This is similar to the case of an obstructed atomic insulator, as presented
in Ref. \cite{Hilberbasis_PRB2025}.}

It is also interesting to mention the possible 
future routes for orbital-obstructed topology. 
As we briefly mentioned in the previous section,
fragile topology can be present in an indirect form for the materials with a
half-filled EBR from $2d$ WP \cite{ArroyoGascon2023}. The role that this
kind of topology plays in the boundary responses is of great interest
since there are few cases of electronic systems where fragility leads to
measurable phenomena. 

Another appealing research line is related to correlation effects in these
systems. In particular, many of the materials reported here are known to
realize the so-called charge-transfer insulating state 
\cite{khomskii-book,Ushakov_2011}, which entails 
states where the low energy excitations of the material are related to an
unexpected electron transfer from the X atoms to the M atoms. This process
is well-established for example in bulk three-dimensional transition metal
oxides such as NiO \cite{NiO_PRL07}, but it is less studied in layered
transition metal dichalcogenide systems and other samples where M is an alkali
metal like Mg, or a group IV element like Ge or Sn.
Therefore, it would be intriguing to characterize how the orbital obstruction is
affected when considering electronic correlation effects. 

\vspace{1cm}

\section*{Acknowledgements}
We acknowledge the
financial support of the Agencia Estatal de Investigación of Spain under grant
PID2022-136285NB-C31 and from the Comunidad de Madrid through the
(MAD2D-CM)-UCM5 project, funded by the Recovery, Transformation and Resilience
Plan, and by NextGenerationEU from the European Union. O.A.G acknowledges
funding from European Union NextGenerationEU/PRTR project Consolidación
Investigadora CNS2022-136025. S.B. acknowledges the support of the Postdoctoral
Grant from the Universidad Técnica Federico Santa María. M. P. acknowledges the
financial support of Chilean FONDECYT by grant 1211913. S.B acknowledges the support
of the postdoctoral position at Universidad Técnica Federico Santa María. Finally,
we thank the Centro de Supercomputación de Galicia, CESGA, (www.cesga.es, Santiago de
Compostela, Spain) and Supercomputación Castilla y León (SCAYLE) for providing
access to their supercomputing facilities.

\section*{Appendix} 

\subsection*{Computational details}
Electronic structure calculations for all materials were carried out by means of the
Quantum ESPRESSO first-principles code
\cite{giannozzi_quantum_2009,giannozzi_advanced_2017} 
using the generalized gradient approximation (GGA) and
Perdew–Burke-Ernzerhof (PBE) exchange-correlation functional. Spin-orbit
coupling was considered throughout the self-consistent calculations in all
cases. A kinetic energy wavefunction cutoff of 100 Ry and a $8\times 8\times 1$
Monkhorst-Pack reciprocal space grid were employed for all calculations. All structures were relaxed
until the forces were under 0.001 eV/\AA. The initial crystal structures
were obtained from the Computational 2D Materials Database (C2DB) 
\cite{haastrup_computational_2018, gjerding_recent_2021} 
and checked against the Topological Materials Database 
\cite{vergniory_complete_2019,vergniory_all_2022,bradlyn_topological_2017-1}. Afterwards, the irreducible representations were obtained 
using the IrRep code
\cite{iraola_irrep_2022}. 
For the flake calculations, the SIESTA code
\cite{soler_siesta_2002-2,garcia_siesta_2020} 
was used, considering the same
GGA-PBE approximation and spin-orbit coupling. The flake and ribbon numerical calculations
based on the tight-binding model were carried using the Kwant package
\cite{KWANT_2014}.

\subsection*{Tight-binding model functions}

\begin{equation*}
F( r_{\alpha } ,r_{\beta }) =\sum _{m\in \{k_{x} ,k_{y} ,k_{x} +k_{y}\}}
r_{\alpha } \cos m +ir_{\beta } \sin m,
\end{equation*}
where $\alpha$ and $\beta$ $\in \{1,2,4\}$.

\begin{equation*}
G( r_{3}) =\frac{2}{\sqrt{3}} r_{3}\left[ i\sin k_{x} +\omega ^{\ast }
\sin k_{y} +\omega \sin( k_{x} +k_{y})\right].
\end{equation*}
where $\omega=e^{i\pi/6}$.

\begin{equation*}
\begin{aligned}
L( \xi ) & =\xi \left( z_{1}^{\ast } +z_{2} +z_{3}\right).\\
M( \xi ) & =\xi \left( \phi ^{\ast } z_{1}^{\ast } +\phi z_{2} -z_{3}\right).\\
\tilde{M}( \xi ) & =-\xi \left( \phi z_{1}^{\ast } +\phi ^{\ast } z_{2} +z_{3}\right).\\
\tilde{T}( \xi ) & =-\xi \left( \phi ^{\ast } z_{1} +\phi z_{2}^{\ast } -z_{3}^{\ast }\right).\\
T( \xi ) & =\xi \left( \phi z_{1} +\phi ^{\ast } z_{2}^{\ast } -z_{3}^{\ast }\right).
\end{aligned}
\end{equation*}
Here, $\phi=e^{i\pi/3}$ and 
\begin{equation*}
\begin{aligned}
z_{1} & =e^{i/3( 2k_{x} +k_{y})},\\
z_{2} & =e^{i/3( k_{x} -k_{y})},\\
z_{3} & =e^{i/3( k_{x} +2k_{y})}.
\end{aligned}
\end{equation*}

\subsection*{EBR matrix for space group P$\bar{3}m1$ monolayers}
As stated in Section IV, we perform a Topological Quantum Chemistry analysis of the valence bands of three different monolayers belonging to space group P$\bar{3}$m1. The EBR matrix, which is the same for all P$\bar{3}$m1 monolayers, is detailed in the following.
\begin{equation}
\left(
\begin{array}{cccccccc}
 1 & 0 & 0 & 0 & 1 & 0 & 1 & 0 \\
 0 & 1 & 0 & 0 & 1 & 0 & 0 & 1 \\
 0 & 0 & 1 & 0 & 0 & 1 & 2 & 0 \\
 0 & 0 & 0 & 1 & 0 & 1 & 0 & 2 \\
 1 & 0 & 1 & 0 & 1 & 1 & 1 & 2 \\
 0 & 1 & 0 & 1 & 1 & 1 & 2 & 1 \\
 1 & 1 & 0 & 0 & 0 & 1 & 1 & 1 \\
 0 & 0 & 1 & 1 & 2 & 1 & 2 & 2 \\
\end{array}
\right).
\end{equation}
Correspondingly, the $\Delta$ matrix that obeys $\Delta = L\cdot EBR \cdot R$ is stated below. More information on the EBR matrix can be found in Section IV, Eqs. 6 - 9. 
\begin{equation}
\left(
\begin{array}{cccccccc}
 1 & 0 & 0 & 0 & 0 & 0 & 0 & 0 \\
 0 & 1 & 0 & 0 & 0 & 0 & 0 & 0 \\
 0 & 0 & 1 & 0 & 0 & 0 & 0 & 0 \\
 0 & 0 & 0 & 1 & 0 & 0 & 0 & 0 \\
 0 & 0 & 0 & 0 & 1 & 0 & 0 & 0 \\
 0 & 0 & 0 & 0 & 0 & 2 & 0 & 0 \\
 0 & 0 & 0 & 0 & 0 & 0 & 0 & 0 \\
 0 & 0 & 0 & 0 & 0 & 0 & 0 & 0 \\
\end{array}
\right).
\end{equation}


%

\end{document}